\documentclass[twocolumn,superscriptaddress,letter]{revtex4}%
\usepackage{amssymb}
\usepackage{color}
\usepackage{graphicx}
\usepackage{dcolumn}
\usepackage{bm}
\usepackage[header,title,page,titletoc]{appendix}
\usepackage{amsmath}
\usepackage{amsfonts}%
\providecommand{\U}[1]{\protect \rule{.1in}{.1in}}

\begin{document}
\title{Effects of Attractive correlation on Topological Flat-bands Model}
\author{Chun-Li Zang}
\thanks{zangzys@mail.bnu.edu.cn}
\affiliation{Department of Physics, Beijing Normal University, Beijing, 100875, P. R. China}
\author{Jing He}
\affiliation{Department of Physics, Hebei Normal University, Hebei, 050024, P. R. China}
\author{Ya-Jie Wu}
\affiliation{Department of Physics, Beijing Normal University, Beijing, 100875, P. R. China}

\begin{abstract}
In this paper, we study the effects of attractive correlation on the
topological insulator ($TI$) with topological flat-bands using
 an extended attractive Kane-Mele-Hubbard model (KMHM).
 In the KMHM, we found a quantum phase transition from $TI$ to the
superconductor ($SC$) state upon the increasing of the attractive Hubbard
interaction $U$ at the mean field level. This type of $SC$ phase transition is
different from the traditional $SC$ phase transition which develops from the
gapless Fermi Liquid. Cooperon-type gapped excitations exist in the $TI$ side
near this type of $SC$ phase transition.

\end{abstract}
\maketitle

\section{Introduction}

The integer quantum Hall (IQH) effect was first observed in a two dimensional
(2D) electron gas subjected to strong perpendicular magnetic field\cite{von}.
This effect provides the first example of topological states that beyond the
Landau symmetry breaking paradigm. After this observation, Haldane, in 1988,
proposed a model (the Haldane model) and he found a state which also has the
IQH effect in this model but the realization of this state doesn't need the
external magnetic field\cite{haldane}. The Haldane model describe a system
spinless fermions and the time reversal symmetry is broken in this model by
its complex NNN hoppings. The state that Haldane found is also a topological
one. The two type topological states we mentioned above that support IQH
effect could be characterized by an topological invariant - TKNN number (the
Chern number)\cite{thouless}. Following the lesson from the Haldane model,
people wonder naturally that whether the fractional quantum Hall (FQH) effect
could also be realized in a model without external magnetic field. Recently,
models with topologically nontrivial flat-bands (TFBs) were found to be an
promising candidate to realize FQH effect without external magnetic
field\cite{flat,flat2,flat3}.

Along with the IQH and FQH states in which time-reversal symmetry breaking is
required, time-reversal symmetry protected topological states of matters are
also discovered in the quantized spin Hall effect (QSH)\cite{km,berg}. People
called them $Z_{2}$ topological insulator ($TI$) state. The typical model for
the $Z_{2}$ topological insulator is the Kane-Mele (KM) model\cite{km}.
Recently, the correlated effects in $TI$ states are studied by various groups
with the Kane-Mele-Hubbard model as the starting point\cite{fchzhang,dung}.

In this paper, we investigate the effects of attractive interaction to a TFBs
system using an attractive Kane-Mele-Hubbard model on the honeycomb lattice:
$H_{\mathrm{KMH}}$. There are two important parts in $H_{\mathrm{KMH}}$ :
$H_{\mathrm{EKM}}$ and $H_{U}$, see Eq.(\ref{kmh}). In the original KM model,
the authors generalizes Haldane's model \cite{haldane} to include spin with
time reversal invariant spin-orbit interactions\cite{km}. So the KM model is a
free model. $H_{\mathrm{EKM}}$ is the free limit ($U=0$) of $H_{\mathrm{KMH}}$
and is a general case of KM model i.e., the next-nearest-neighbor (NNN)
hoppings for the spin-$\uparrow$ and spin-$\downarrow$ electrons are complex
valued and complex conjugate to each other in $H_{\mathrm{EKM}}$. So the
Kane-Mele model is a special case of $H_{\mathrm{EKM}}$ i.e., when the complex
valued NNN hoppings is purely imaginary or the Haldane phase $\hat{\phi}_{ij}$
in Eq.(\ref{ekm}) is $\pi/2$. There are particle-hole symmetry in the
half-filing orignal Kane-Mele model i.e., when the complex valued NNN hoppings
is purely imaginary it reduce to a spin-orbit interactions\cite{ph}. So there
are no particle-hole symmetry in $H_{\mathrm{EKM}}$ except $\hat{\phi}_{ij}=$
$\pi/2$. If we varying the Haldane phase $\hat{\phi}_{ij}$ there exist
a so called TFBs limit in $H_{\mathrm{EKM}}$ \cite{flat,flat2,flat3}, see
Fig.\ref{Fig.1}(b).

Generally for the system with flat-bands, the kinetic energy will be quite
suppressed and the interaction becomes highly relevant. In this paper, $H_{U}$
in $H_{\mathrm{KMH}}$ is treated by self consisted mean field method, at this level we
find a $TI-SC$ quantum phase transition as the Hubbard interaction strength
increases beyond a critical value $U_{c}$ in the $H_{\mathrm{KMH}}$ model, see
Fig.\ref{Fig.4}. In this type $SC$ phase transition, there is a kind of
Cooperon-type excitations in the insulator side near the $SC$ phase transition
and the Cooperon-type excitations are gapped before its condensation
\cite{cooperon,cooperon1}.

\section{Topological Flat-band model}

The Hamiltonian that we study in this paper is $H_{\mathrm{KMH}}$, the
attractive Kane-Mele-Hubbard model which can be writed as:
\begin{equation}
H_{\mathrm{KMH}}=H_{\mathrm{EKM}}+H_{U}-\mu \sum_{i,\sigma}\hat{c}_{i\sigma
}^{\dagger}\hat{c}_{i\sigma},
\label{kmh}%
\end{equation}
where $\sigma=\uparrow$, $\downarrow$ denotes the spin degree freedom,
$H_{U}=-U\sum_{i}\hat{n}_{i\uparrow}\hat{n}_{i\downarrow}$ represent the
on-site Hubbard type attractive interaction, $\mu$ is the chemical potential.
We call $H_{\mathrm{EKM}}$ the extended KM model whose Hamiltonian is written as:
\begin{align}
H_{\mathrm{EKM}}  &  =-t\sum_{\left \langle i,j\right \rangle ,\sigma}\hat
{c}_{i\sigma}^{\dagger}\hat{c}_{j\sigma}-t^{\prime}\sum_{\left \langle
\left \langle i,j\right \rangle \right \rangle ,\sigma}e^{i\hat{\phi}_{ij}%
\hat{\sigma}_{z}}\hat{c}_{i\sigma}^{\dagger}\hat{c}_{j\sigma}\label{ekm}\\
&  -t^{\prime \prime}\sum_{\left \langle \left \langle \left \langle
i,j\right \rangle \right \rangle \right \rangle ,\sigma}\hat{c}_{i\sigma
}^{\dagger}\hat{c}_{j\sigma}+h.c.\nonumber
\end{align}

In $H_{\mathrm{EKM}}$, the first term describes the nearest-neighbor (NN)
hopping on the honeycomb lattice and its hopping strength $t$ is set as
the unit of energy in the rest of this paper.
The second term describes the next-nearest-neighbor (NNN) hopping with a
Haldane type complex strength $t^{\prime}e^{i\hat{\phi}_{ij}}$,
the phase factor in this term is spin dependended, i.e.,
$e^{i\hat{\phi}_{ij}}=e^{i\phi}$ for spin $\uparrow$ electron that hopping
clockwise in the fundamental plaquette as shown by the blue arrow in Fig.1(a) and $e^{i\hat
{\phi}_{ij}}=e^{-i\phi}$ for spin $\downarrow$ electron.
This term recover the original spin-orbit
coupling term of KM model when $\phi=\pi/2$
(so here $t^{\prime}$ cannot be regarded as the the
spin-orbit interaction strength).
This term also reduces $H_{\mathrm{EKM}}$'s
full spin rotational SU(2) symmetry to a U(1) symmetry.
Thus there is time reversal symmetry (TRS) but
no full spin rotation symmetry in $H_{\mathrm{EKM}}$.
The last term represents the next-next-nearest-neighbor (NNNN) hopping with
strength $t^{\prime \prime}$, the energy bands of $H_{\mathrm{EKM}}$
 could achieve a flat-bands limit with this term in it
 (more about this limit see below). In
this paper, we consider the half-filling case of $H_{\mathrm{KMH}}$ i.e.,
the chemical potential $\mu=0$.

\begin{figure}[ptbh]
\centering
\includegraphics[width=0.18\textwidth]{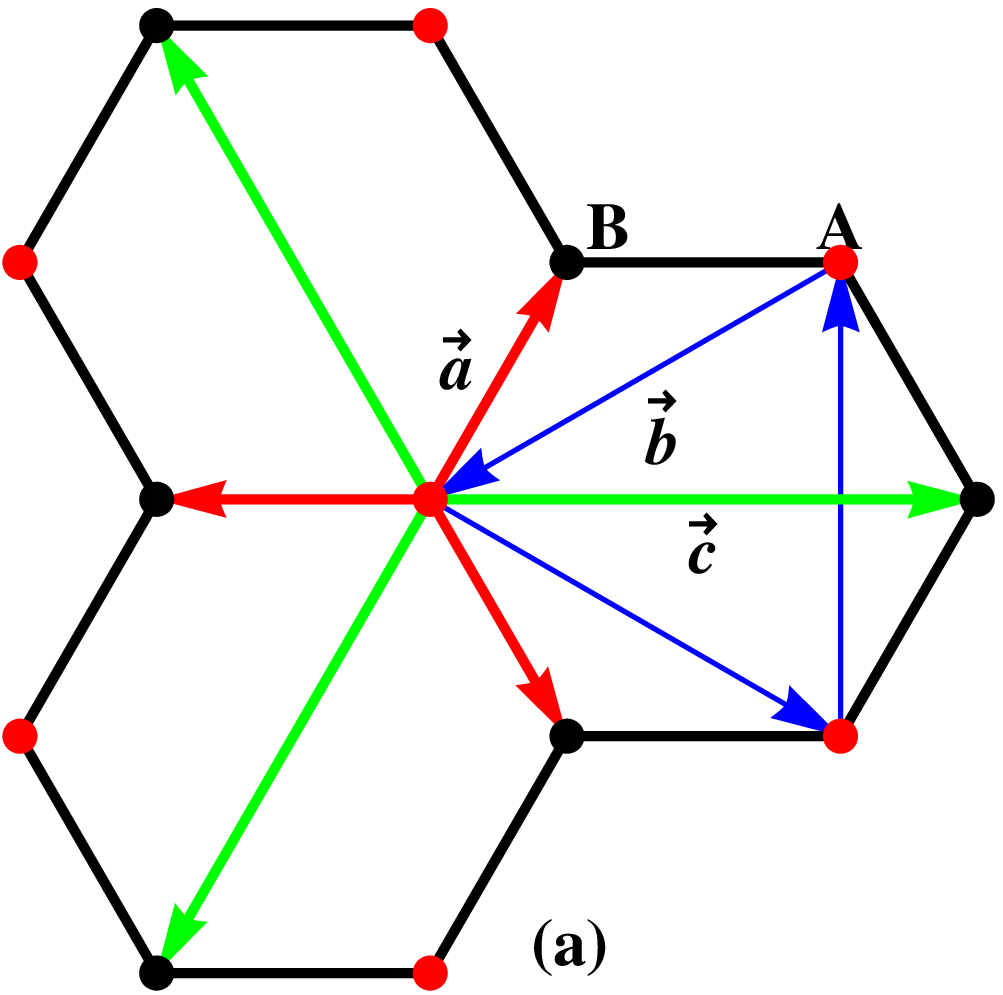}
\includegraphics[width=0.29\textwidth]{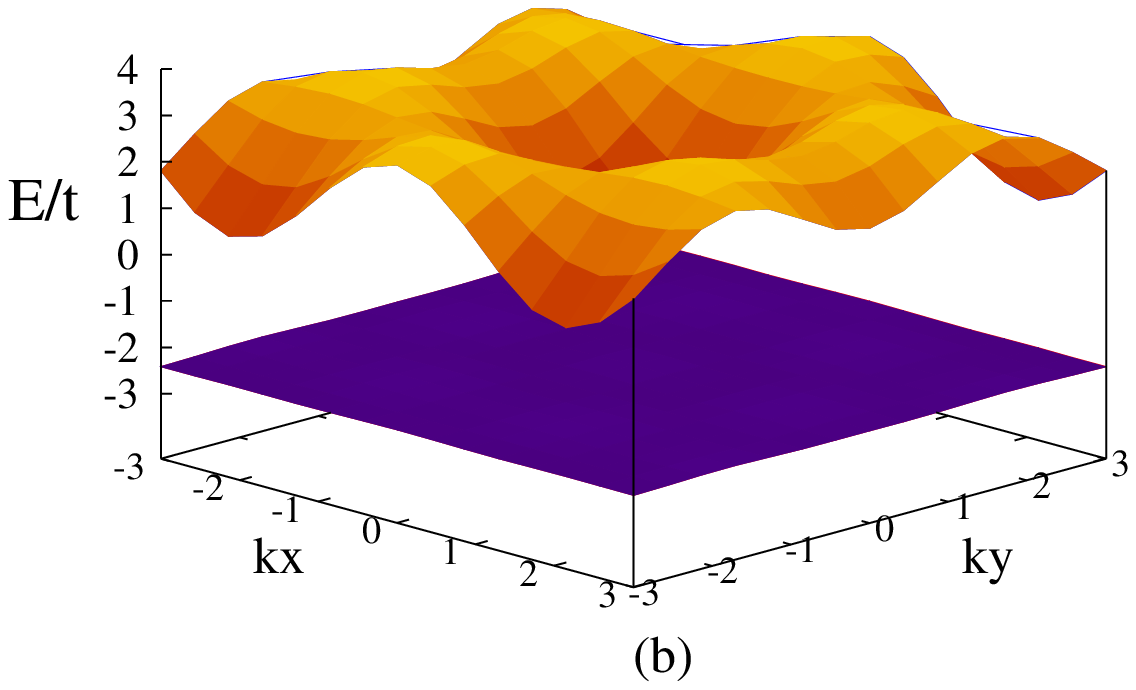}
\caption{(color online)(\textbf{a}) The honeycomb lattice consists of two sublattices: A (the
red sites) and B (the black sites). The NN (red), NNN (blue) and NNNN (green)
displacement vectors are denoted by $\vec{a}$, $\vec{b}$ and $\vec{c}$
respectively. The arrow in $\vec{b}$ also denote the positive phase direction
of the complex factor $e^{i\phi_{ij}}$ that company the NNN hopping.
(\textbf{b}) The TFBs in the energy spectrum of the $H_{\mathrm{EKM}}$
with parameters: $t^{^{\prime}}/t=0.6$, $t^{^{\prime
\prime}}/t=-0.58$, $\phi_{ij}=0.4\pi$.}%
\label{Fig.1}%
\end{figure}

The $H_{\mathrm{EKM}}$ model is the free limit ($U=0$) of the $H_{\mathrm{KMH}%
}$ model. In terms of the basis vector $\Phi^{\dagger}\left(k\right)=\left[
a_{k\uparrow}^{\dagger},b_{k\uparrow}^{\dagger},a_{k\downarrow}^{\dagger
},b_{k\downarrow}^{\dagger}\right]$, $H_{\mathrm{KMH}}$ could be expressed in a
block-diagonal matrix form as: $H_{\mathrm{EKM}}=\sum_{k}\Phi^{\dagger}\left(
k\right) H_{0}\left(k\right)\Phi\left(k\right)$
where
\begin{align}
H_{0}(k)  =I_{0}C+(
\begin{array}
[c]{cc}%
h_{\uparrow}\cdot \sigma & 0\\
0 & h_{\downarrow}\cdot \sigma
\end{array}
)
\label{ekmk}
\end{align}
with $\mathbf{h}_{\uparrow}=\left(  h^{x},h^{y},h^{z}\right)  $,$\  \mathbf{h}%
_{\downarrow}=\left(  h^{x},h^{y},-h^{z}\right)  $, $I_{0}$ is the $4\times4$
unit matrice and $\sigma$ is the Pauli matrice that act on the A, B
sublattice space of the bipartite honeycomb lattice. The three components of
vector $\mathbf{h}_{\uparrow \left(  \downarrow \right)  }$ $=(\operatorname{Re}%
\gamma \left(  k\right)  ,-\operatorname{Im}\gamma \left(  k\right)  ,D)$, where
$\gamma \left(  k\right)  =\xi \left(  k\right)  +\xi^{^{\prime \prime}}\left(
k\right)  $ with $\xi \left(  k\right)  =-t\sum_{i=1}^{3}e^{ik\cdot \vec{a}_{i}%
}$ and $\xi^{\prime \prime}\left(  k\right)  =-t^{\prime \prime}\sum_{i=1}%
^{3}e^{ik\cdot \vec{c}_{i}}$, with $\vec{a}_{1}=\left(  1/2,-\sqrt{3}/2\right)
$, $\vec{a}_{2}=\left(  1/2,\sqrt{3}/2\right)  $, $\vec{a}_{3}=\left(
-1,0\right)  $.
\begin{align}
C  &  =-2t^{^{\prime}}\cos \phi_{ij}\sum_{i=1}^{3}\cos[k\cdot \vec{b}_{i}] \nonumber \\
D  &  =-2t^{^{\prime}}\sin \phi_{ij}\sum_{i=1}^{3}\sin[k\cdot \vec{b}_{i}]
\label{cd}
\end{align}
with $\vec{b}_{1}=\left(  3/2,-\sqrt{3}/2\right)  $, $\vec{b}_{2}=\left(
0,\sqrt{3}\right)  $, $\vec{b}_{3}=\left(  -3/2,-\sqrt{3}/2\right)$, see
Fig.\ref{Fig.1}(a). We set the lattice constant $a=1$ in the rest of the
paper. The energy dispersion of the $H_{\mathrm{EKM}}$ can be founded by
diagonalize $H_{0}\left(k\right)$:%

\begin{equation}
E_{n}\left(  k\right)  =C\pm \sqrt{\left(  h^{x}\right)  ^{2}+\left(
h^{y}\right)  ^{2}+\left(  h^{z}\right)  ^{2}}.
 \label{freesp}%
\end{equation}

There are two energy bands in $E_{n}(k)$, both of them are doubly degenerate
due to the spin degree freedom and have a flat-bands limit, i.e., with parameters:
$\phi=0.4\pi$, $t^{^{\prime}}=0.6t$, $t^{^{^{\prime \prime}}}=-0.58t$
\cite{flat2}. In this limit, the lower bands of $E_{n}(k)$
will become flat, see Fig.\ref{Fig.1}(b). The flatness ratio of the flat-bands
in this limit (the ratio of the band gap over bandwidth) can reach about $50$.
we know that the NN and NNNN hopping energy $\xi(k)$ and $\xi
^{\prime \prime}(k)$ are vanish at the two Dirac points:
$(2\pi/3,2\pi/3\sqrt{3})$ and $(2\pi/3,-2\pi/3\sqrt
{3})$ in momentum space, but the NNN hopping energy $\xi
_{a(b)}^{\prime}(k)$ dose not, just like the spin-orbit coupling term in KM model,
it will opens an energy gap at those two Dirac points. So $H_{\mathrm{KMH}}$ is gapped in its
free limit. We denote this energy gap as $\Delta_{t^{\prime}}$, it's the bulk gap of the $TI$ state
since before $H_{\mathrm{KMH}}$ model entering the $SC$ phase it is in a $TI$ phase, the magnitude of
$\Delta_{t^{\prime}}$ is related to $t^{^{\prime}}$.
Thus we called $\Delta_{t^{\prime}}$ the bulk gap which playing the same role as the gap of semiconductor in \cite{cooperon}.

In $H_{\mathrm{KMH}}$'s flatband limit, the Chern number for the spin up and down electron that
filled the two lowest bands of $E_{n}(k)$ can be calculate in this way:
\[
C_{\uparrow/\downarrow}=\frac{1}{4\pi}\int \mathbf{n}\cdot \left(
\frac{\partial \mathbf{n}}{\partial k_{x}}\times \frac{\partial \mathbf{n}%
}{\partial k_{y}}\right)  d^{2}k
\]
The Chern number of spin $\uparrow$ components electron is $C_{\uparrow}=1$
with $\mathbf{n=}$ $\mathbf{h}_{\uparrow}/\left \vert \mathbf{h}_{\uparrow
}\right \vert $ and $C_{\downarrow}=-1$ with $\mathbf{n=}$ $\mathbf{h}%
_{\downarrow}/\left \vert \mathbf{h}_{\downarrow}\right \vert $, reflecting the
TRS in $H_{\mathrm{EKM}}$. The spin Chern number
$C_{s}=\left(  C_{\uparrow}-C_{\downarrow}\right)  /2=1$ reflect the QSH in
the free limit of $H_{\mathrm{KMH}}$ model\cite{meng}.

\begin{figure}[ptb]
\includegraphics[width=0.45\textwidth]{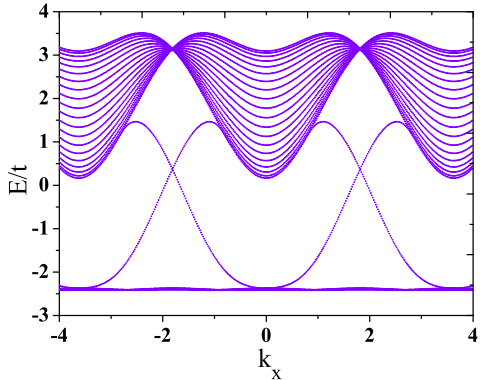}
\caption{(color online) The
edge states of $H_{\mathrm{EKM}}$ in its flat-bands
limit in a cylinder geometry with a zigzag edge. This
is a typical edge states of TRS topological insulator which also
reflecting the nontrivial topology of the flat-bands.}%
\label{Fig.2}%
\end{figure}

This conclusion could be further verified by the presence of edge states in the
energy spectrum of $H_{\mathrm{EKM}}$ when it's imposed a cylinder boundary
condition with zigzag edges, i.e., we set periodic boundary condition along
the system's $x$-direction and open boundary condition along the
$y$-direction. The numerical results are depicted in Fig.\ref{Fig.2}, from
which we can see that there are topologically protected edge states, which is
one of the signatures of topological states of matter. So in half-filling and
flat-bands limit, with the two lowest flat-bands are filled, $H_{\mathrm{EKM}%
}$ support the so called TFBs. Thus in the flat-bands limit of $H_{\mathrm{EKM}}$ model,
it is a $Z_{2}$ $TI$ with TFBs as a consequence of the TRS in $H_{\mathrm{EKM}}$.

\begin{figure}[ptb]
\includegraphics[width=0.45\textwidth]{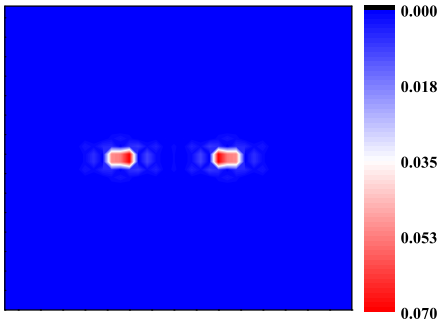}
\caption{(color oline) The
contour plots of the particle density distribution of the wave function of
the zero-energy modes in coordinate space, this pair of zero-energy modes are bound to
the $\pi$-fluxes.
The values of particle density
are indicated by color in this figure,the parameters are in $H_{\mathrm{KMH}}$'s
flat-bands limit.}%
\label{Fig.3}%
\end{figure}

On the other hand, the response of $TI$ to topological defects such as: $\pi
$-fluxes has also been suggested as a probe of the nontrivial topology of the $Z_{2}$ $TI$ \cite{assaad}.
Through numerical calculation, we really found that there are zero-energy modes in the energy spectrum
if a pair of $\pi$-fluxes are adiabatically inserted into the $H_{\mathrm{EKM}}$ model
in its flat-bands limit.
From the particle density distribution in Fig.\ref{Fig.3},
we can see that those zero-energy modes are really located around the $\pi$-fluxes in coordinate space.
Those results imply that the $H_{\mathrm{EKM}}$ model's topological non
trivial in its flat-bands limit.

\section{Interacting flat-band model}

\begin{figure}[ptb]
\includegraphics[width=0.23\textwidth]{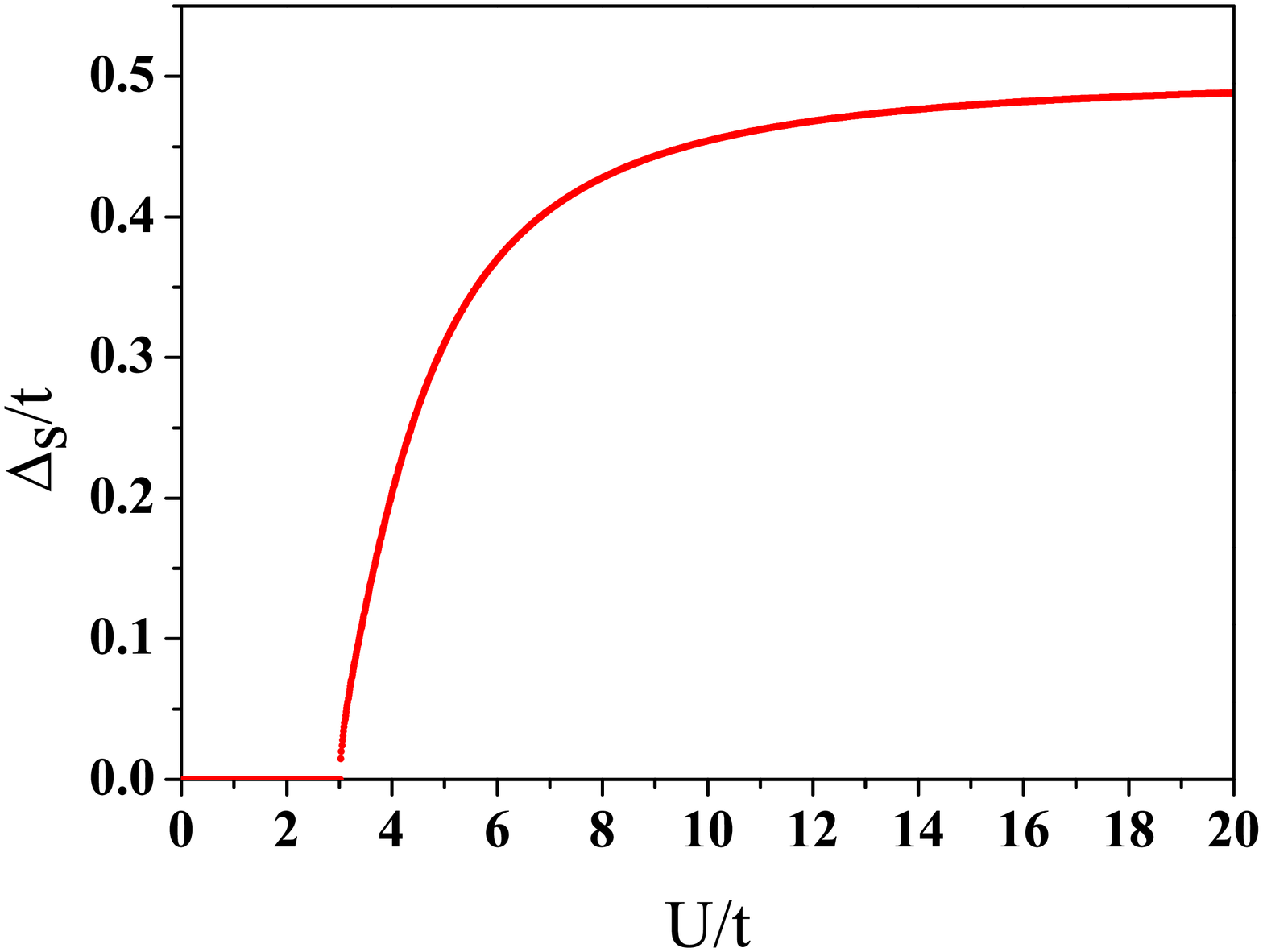}
\includegraphics[width=0.23\textwidth]{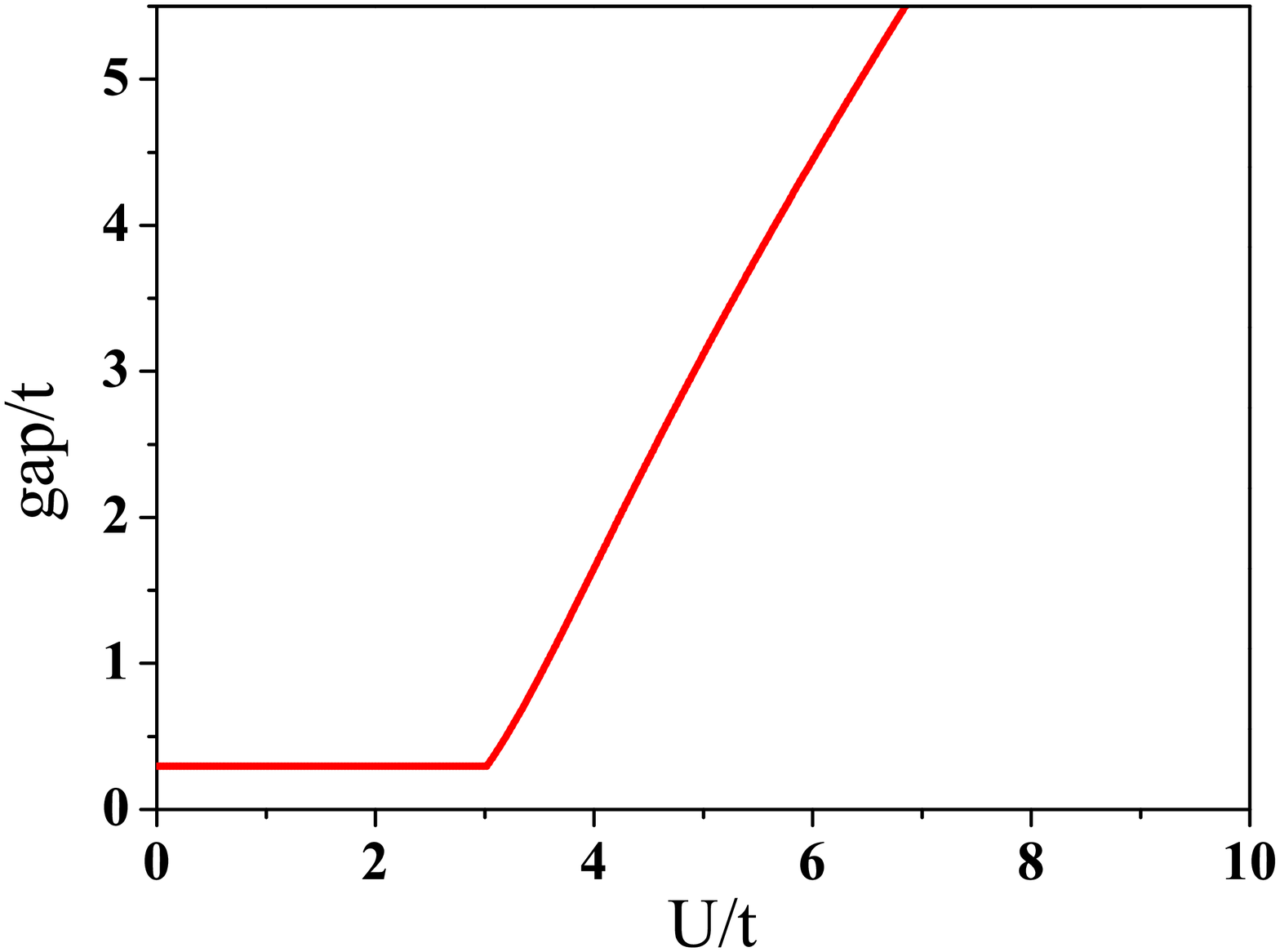}
\caption{(color online)
(\textbf{a}) The mean field results of $\Delta_{s}$ we
solve from Eq.(\ref{self}) at $H_{\mathrm{KMH}}$'s flat-bands limit.
(\textbf{b}) The excitation energy gap of the $SC$ phase as a function of the interaction strength $U$,
this excitation gap increases
with $U$ monotonically so there is no further topological phase transition upon varying $U$.}%
\label{Fig.4}%
\end{figure}

In this section, we study the effects of attractive Hubbard type interaction
of $H_{\mathrm{KMH}}$ model in the flat-bands limit of $H_{\mathrm{EKM}}$ with
the mean field method. In this limit, the resulting mean field phase diagram
of the $H_{\mathrm{KMH}}$ model could be see from Fig.\ref{Fig.4}(a). From
Fig.\ref{Fig.4}(a) we could see that the $TI$ phase in the $H_{\mathrm{KMH}}%
$'s free limit is unstable against an SC phase transition as the
interaction strength increase beyond a finite critical\ value: $U_{c}%
\simeq3.03t$.

According to the self-consistent mean field method, firstly we
introduce an $SC$ order parameter fields to decouple the Hubbard term $H_{U}$
in $H_{\mathrm{KMH}}$. The $SC$ order parameter fields are defined as:
$\Delta_{s}=\left \langle \hat{c}_{i\uparrow}\hat{c}_{i\downarrow}\right \rangle
$. By the mean field approximation, we substitute
$\hat{c}_{i\uparrow}^{\dag}\hat{c}_{i\downarrow}^{\dag}=\left \langle \hat
{c}_{i\uparrow}^{\dag}\hat{c}_{i\downarrow}^{\dag}\right \rangle +\delta^{\dag
}$ and $\hat{c}_{i\uparrow}\hat{c}_{i\downarrow}=\left \langle \hat
{c}_{i\uparrow}\hat{c}_{j\downarrow}\right \rangle +\delta$ into the Hubbard
term $H_{U}$ in Eq.(\ref{kmh}), with $\delta^{\dag}=\left(  \hat{c}%
_{i\uparrow}^{\dag}\hat{c}_{i\downarrow}^{\dag}-\left \langle \hat
{c}_{i\uparrow}^{\dag}\hat{c}_{i\downarrow}^{\dag}\right \rangle \right)  $ and
$\delta=\left(  \hat{c}_{i\uparrow}\hat{c}_{i\downarrow}-\left \langle \hat
{c}_{i\uparrow}\hat{c}_{i\downarrow}\right \rangle \right)  $ two small
quantities. We discard the second order terms of $\delta^{\dag}$ and
$\delta$, then $H_{U}$ could be decoupled into a bilinear form of $\hat{c}%
_{i\sigma}^{\dag}$ and $\hat{c}_{i\sigma}$ as:
\begin{align}
H_{U}= & -U\sum_{i}\hat{n}_{i\uparrow}\hat{n}_{i\downarrow}\simeq \nonumber \\
& -U\sum_{i}\left[\Delta_{s}^{\dagger}\hat{c}_{i\downarrow}\hat
{c}_{i\uparrow}+\Delta_{s}\hat{c}_{i\uparrow}^{\dag}\hat{c}_{i\downarrow
}^{\dag}\right]+UN_{s}\left \vert \Delta_{s}\right \vert ^{2}\text{.}
\label{hubbard}
\end{align}

The honeycomb lattice's bipartite and translational invariant, so we could introduce
the usual Fourier transformation to the electron creation (destruction)
operator $\hat{c}_{i\sigma}^{\dag}$ $(\hat{c}_{i\sigma})$ in
$H_{\mathrm{KMH}}$ and we denote the
fourier transform of $\hat{c}_{i\in A,\sigma}^{\dag}$ and $\hat{c}_{i\in
B,\sigma}^{\dag}$ as $\hat{a}_{k\sigma}^{\dag}$ and $\hat{b}_{k\sigma
}^{\dagger}$ respectively:%

\begin{align}
\hat{c}_{i\in A,\sigma}^{\dag} & =\frac{1}{\sqrt{N_{s}}}\sum_{k}\hat{a}%
_{k,\sigma}^{\dag}e^{ik\cdot R_{i}}\text{,} \nonumber \\
\hat{c}_{i\in B,\sigma}^{\dag} & =\frac{1}{\sqrt{N_{s}}}\sum_{k}\hat{b}_{k,\sigma}^{\dag}e^{ik\cdot R_{i}}\text{.}
\label{ft}%
\end{align}
where $N_{s}$ is the number of unite cells and $k$ belong to the first
Brillouin Zone of the honeycomb lattice.
Now we substitute Eq.(\ref{hubbard}) and Eq.(\ref{ft}) into Eq.(\ref{kmh})
we could got the momentum space form of $H_{\mathrm{KMH}}$: $H_{\mathrm{KMH}}(k)$.
In the Nambu basis:
$\psi^{\dagger}(k)=(\hat{a}_{k\uparrow}^{\dag}, \hat{a}_{-k\downarrow},\hat{b}_{k\uparrow
}^{\dag}, \hat{b}_{-k\downarrow})$, $H_{\mathrm{KMH}}(k)$ could be casted into a matrix form in momentum
space, that's:
\begin{equation}
H_{\mathrm{KMH}}(k)=\sum \limits_{k}\psi^{\dagger}\left(  k\right)  h\left(
k\right)  \psi \left(  k\right)  +UN_{s}\left \vert \Delta_{s}\right \vert ^{2}%
\end{equation}
with the $4\times4$ matrix $h\left(  k\right)  $:
\begin{equation}
h\left(  k\right)  =\left(
\begin{array}
[c]{cccc}%
C+D & -U\Delta_{s} & \gamma \left(  k\right)  & 0\\
-U\Delta_{s} & -C-D & 0 & -\gamma \left(  k\right) \\
\gamma^{\dagger}\left(  k\right)  & 0 & C-D & -U\Delta_{s}\\
0 & -\gamma^{\dagger}\left(  k\right)  & -U\Delta_{s} & -C+D
\end{array}
\right)  \label{hamk}%
\end{equation}
where $C$ and $D$ are the same as in Eq.(\ref{cd}). The quasi-particles
spectrum can be found by diagonalize the Hamiltonian(\ref{hamk}) in the
momentum space:
\begin{equation}
E_{n}\left(  k\right)  =\pm \sqrt{e_{1}\pm2e_{2}}.
\label{spectrum}%
\end{equation}

where%

\begin{align*}
e_{1}  &  =C^{2}+D^{2}+\left \vert \gamma \left(  k\right)  \right \vert
^{2}+\left(  U\Delta_{s}\right)  ^{2},\\
e_{2}  &  =\sqrt{C^{2}\left(  D^{2}+\left \vert \gamma \left(  k\right)
\right \vert ^{2}\right)  }.
\end{align*}

We would investigate the instability of the $TI$ phase of $H_{\mathrm{KMH}}$
in the presence $SC$ fluctuation as the interaction strength $U$ increases by
minimizing the ground state's energy $E_{0}=\sum_{n,k} E_{n}(k)$
against the $SC$ order parameters $\Delta_{s}$, i.e.,
$\partial E_{0}/\partial \Delta_{s}=0$. Then the self-consistent mean field
equation is\ given by:
\begin{equation}
1=\frac{U}{2N_{s}}%
{\displaystyle \sum \limits_{k}}
\left(  \frac{1}{E_{1}(k)}+\frac{1}{E_{2}(k)}\right).
 \label{self}%
\end{equation}
where $E_{1}(k)=-\sqrt{e_{1}+2e_{2}}$ and $E_{2}(k)=-\sqrt{e_{1}%
\newline-2e_{2}}$ are the two lowest filled bands in Eq.(\ref{spectrum}).

The mean field solution of the self-consistent equation Eq.(\ref{self})
are plotted in Fig.\ref{Fig.4}(a) from which we can see that beyond a critical
value of the attractive interaction strength: $U_{c}\simeq3.03t$ the $TI$ phase
of the $H_{\mathrm{KMH}}$ model in its flat-bands limit become unstable against a
$SC$ phase transition.

The quasi-particles excitation gap in the $SC$ phase
increases with $U$ monotonically, see Fig.\ref{Fig.4}(b).
This mean that there is no gap closing and no further topological phase transition upon varying $U$.

\begin{figure}[ptb]
\includegraphics[width=0.45\textwidth]{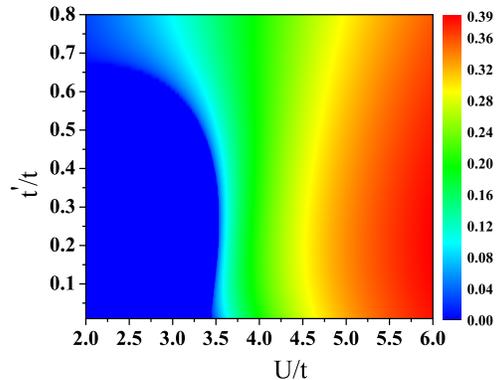}
\caption{(color online) The
mean field results of the $SC$ order parameter $\Delta_{s}$ we solve from Eq.(\ref{self}) over a range of $U$
and $t^{^{\prime}}$, Other parameters are in the flat-bands limit.}%
\label{Fig.5}%
\end{figure}
\begin{figure}[ptb]
\includegraphics[width=0.45\textwidth]{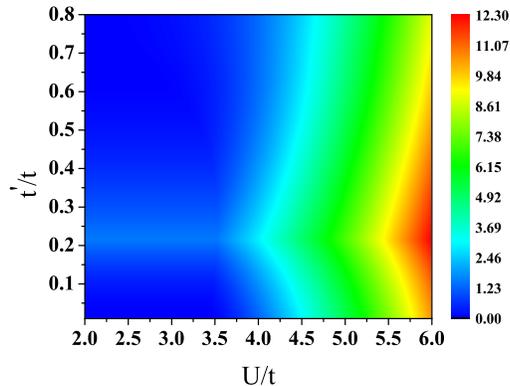}
\caption{(color online) The
mean field results of the excitation energy gap in the $SC$ phase
over a range of $U$ and $t^{^{\prime}}$ while
other parameters are in the flat-bands limit.}%
\label{Fig.6}%
\end{figure}

Fig.\ref{Fig.5} and Fig.\ref{Fig.6} is the mean filed
results of the $SC$ order parameter $\Delta_{s}$ we obtained from
solving Eq.(\ref{self})and the excitation gap of $SC$ phase
over a range of $t^{^{\prime}}$ and $U$ with the
other parameters are fixed in the flat-bands limit.
Note that in the presence of the bulk gap $\Delta_{t^{\prime}}$ the
order parameter $\Delta_{s}$ and the excitation gap of $SC$ phase are not
the same like in the traditional $SC$ phase transition\cite{cooperon}.
From Fig.\ref{Fig.5} we could
infer that the critical interaction strength $U_{c}$ which separate
the $TI$ and $SC$ phase increases nearly linearly with $t^{^{\prime}}$.

\begin{figure}[ptb]
\includegraphics[width=0.45\textwidth]{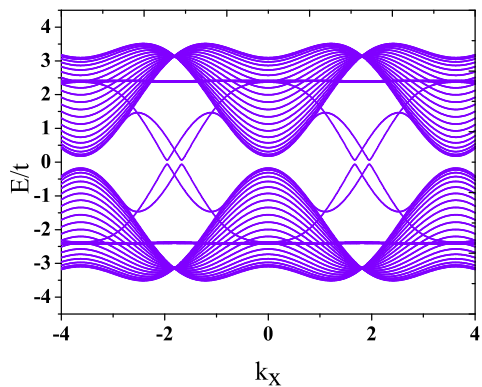}
\caption{(color online) Energy bands of $H_{\mathrm{KMH}}$ in its $SC$ phase
in a cylinder geometry
 with \textquotedblleft zigzag\textquotedblright\ boundary.
 The $SC$ phase is at $U=3.91$ with $\Delta_{s}=0.016t$,
 the other parameters are in the flat-bands limit. }%
\label{Fig.7}%
\end{figure}

\begin{figure}[ptb]
\includegraphics[width=0.45\textwidth]{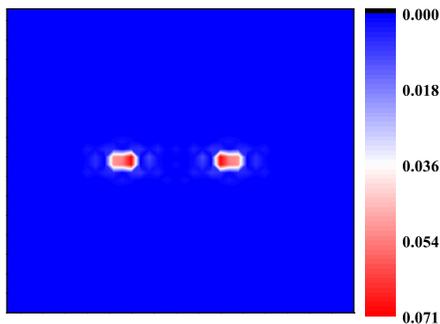}
\caption{(color online) The
contour plots of the particle density distribution of the finite-energy bound states
 in the coordinate space. We can see that the finite-energy states are bound
to the $\pi$-fluxes. The values of particle
density are indicated by color. $H_{\mathrm{KMH}}$'s in its $SC$ phase and
flat-bands limit. }%
\label{Fig.8}%
\end{figure}

As a consequence of TRS in $H_{\mathrm{KMH}}$, the Chern number will be zero
and we verified this by a numerically calculation. We calculate the Chern
number of the two lowest bands of $h\left(  k\right)  $ is $h\left(  k\right)
$\cite{chern,hatsugai}, the numerical results show that the total Chern number
$C=0$ in the $SC$ phase at$U=3.03$ with $\Delta_{s}=0.0147$ while
the other parameters in the flat-bands limit. This mean that the $SC$ phase transition destroy the
TFBs. The topological property of this $SC$ state can also be see from its
edge excitations, so we also calculate the energy bands of $H_{\mathrm{KMH}}$ in
a cylinder geometry. The numerical results are depicted
in Fig.\ref{Fig.7} from which we can see that the edge states are gapped when
the $SC$ order is developed in $H_{\mathrm{KMH}}$. So beyond the critical
interaction strength $U_{c}$, $H_{\mathrm{KMH}}$ entering a topological trivial $SC$
phase. On the other hand, we found a pair finite-energy bound state in the energy spectrum
if a pair of $\pi$-fluxes are inserted into
the $SC$ phase with periodic boundary condition in both $x$ and $y$ direction,
see Fig.\ref{Fig.8}, comparing with the $\pi$-flux-induced zero-energy modes
in Fig.\ref{Fig.3} in the $TI$ phase, show that the $SC$ phase is topological trivial. This
finite energy bound states and the gapped edge states in
Fig.\ref{Fig.7} verified that the $H_{\mathrm{KMH}}$ model will loss its
topology properties upon entering this s-wave $SC$ phase as the attractive interaction
strength increasing.

So the $SC$ phase in $H_{\mathrm{KMH}}$ develops from a fully gapped $TI$
state. This kind of $SC$ phase transition is different from the traditional
$SC$ phase transition which develops from a Fermi liquid in which the low
energy excitations are gapless. If a $SC$ order is develops from an fully
gapped state, in our case an $TI$ state, there will be an kind of gapped
excitations, the so called Cooperon, in the insulator side of this $SC$ phase
transition\cite{cooperon,cooperon1}. This type of $SC$ phase transition's
mechanism colud be understand in this way. There is a competition between
the $SC$ pairing gap $\Delta_{s}$ and the topological energy gap $\Delta_{t^{\prime}}$
upon the varying of interaction strength
$U$. If the topological energy gap
$\Delta_{t^{\prime}}$ is large enough, the Cooperon excitations will have an
energy gap and will not condense, so the $SC$ order parameter is vanishingly
small. On the other hand, if $\Delta_{t^{\prime}}$ is very small or the $SC$
energy gap is large enough (at large $U$), the Cooperon excitations will become
 gapless and condensed, their condensation will leads to the $SC$ phase
transition at last.

\section{Conclusion}

In this paper, we study the influence of attractive correlation on $TI$ using the
$H_{\mathrm{KMH}}$ model via a self-consistent mean field method. In the mean
filed level, we found a $TI-SC$ phase transition upon increasing
the attractive interaction strength $U$ in this model. The
mechanism leading to this $SC$ phase transition is different from the
traditional ones. There is a competition between the topological energy gap
$\Delta_{t^{^{\prime}}}$ and the $SC$ energy gap before the $SC$ phase
transition. This competition is reflect in the gapped Cooperon excitation,
when the $SC$ pairing gap is larger than $\Delta_{t^{^{\prime}}}$, the Cooperon
excitation will condensed and the $SC$ phase transition occurs in the system
eventually. We think those results may help in better understanding the quantum exotic
states in the correlated topological insulator and the pseudogap state in the
high $T_{C}$ cuprate superconductor\cite{cooperon2}.

\textbf{Acknowledgments.} we thank professor Su-Peng Kou for his many
helpful advices. This research is supported by National Basic Research Program
of China (973 Program) under the grant No. 2011CB921803, 2012CB921704 and NFSC
Grant No.11174035.

\end{document}